# Adjustment for Inconsistency in Adaptive Phase 2/3 Designs with Dose Optimization

Cong Chen [a]*, Mo Huang [b]

[a] Biostatistics and Research Decision Sciences, Merck & Co., Inc., Rahway, NJ 07065, USA

*Corresponding author at: MAILSTOP UG-1CD44, 351 North Sumneytown Pike, North Wales, PA 19454, USA. E-mail address: cong_chen@merck.com

[b] Pfizer Inc., Collegeville, PA 19426

**Abstract**

Adaptive Phase 2/3 designs hold great promise in contemporary oncology drug development, especially when limited data from Phase 1 dose-finding is insufficient for identifying an optimal dose. However, there is a general concern about inconsistent results before and after the adaptation. The imperfection in dose selection further complicates the issue. In this paper, we explicitly incorporate the concerns about inconsistency into the statistical analysis under three hypothesis testing strategies. This investigation paves the way for further research in a less explored area.

It is well known that an adaptive Phase 2/3 trial may produce numerically inconsistent results between Stage 1 (Phase 2) and Stage 2 (Phase 3), even when there is no change in the standard of care or eligibility criteria between the two stages [1]. This inconsistency could arise because the trial begins with a limited number of sites in a few countries, which may not be representative of all sites used once recruitment is fully underway. It could occur when the expansion from Phase 2 to Phase 3 is perceived as a positive signal, attracting patients with poor prognoses who normally wouldn't participate in any clinical trial, with unpredictable consequence. It could also simply be due to the difference in follow-up time between Stage 1 and Stage 2 (e.g., a delayed effect may favor Stage 1 while a crossover effect may favor Stage 2). When inconsistency is detected, it can become a serious regulatory review issue and the ability to combine results across independent stages to assess the overall treatment effect becomes questionable [2].

Table 1 illustrates with a hypothetical adaptive Phase 2/3 trial. The study is designed to have 90% power (approximately 510 events) to detect a hazard ratio of 0.75 in overall survival between an experimental arm and a control arm (1:1 randomization ratio) at the 0.025 one-sided significance level. The primary analysis is based on the combined results from the two stages with Stage 1 contributing 30% of the survival information. Without compromising the key message, we assume that the pooled results are tested at the 0.025 level, indicating that an observed hazard ratio of approximately 0.84 would constitute a positive outcome. Regulatory agencies are typically concerned when the overall treatment effect is driven by the Stage 1 data (row 4) while clearly better outcome in Stage 2 (row 1) may also raise questions. In contrast, a lower treatment effect in Stage 1 compared to Stage 2 (row 1) would be a major concern to trial sponsors, as this can reduce the study power and potentially lead to under-estimation of the treatment effect. In conventional adaptive Phase 2/3 designs with dose selection, the under-estimation risk is often low when the dose with best efficacy is carried from Stage 1 to Stage 2, the implicit or explicit assumption under the prevailing statistical approaches to Type I error control such as the p-value combination test [3]. However, it is high in contemporary oncology drug development when the early efficacy endpoint used for dose selection has an uncertain correlation with the primary endpoint as seen in [4] or when safety and tolerability are prioritized over efficacy in dose optimization which makes it uncertain whether the selected dose still has better efficacy [5-6].

Table 1. Survival outcome of a hypothetical adaptive Phase 2/3 survival trial with Stage 1 contributing 30% of the information, where the overall treatment effect is positive at the 0.025 level based on 510 events (90% power for detecting a hazard ratio of 0.75).

| Hazard ratio (overall) | Hazard ratio (Stage 1) | Hazard ratio (Stage 2) | -Log (hazard ratio) (Stage 1/2) | Nominal p-value (Stage 2) |
|---|---|---|---|---|
| 0.84 | 0.954 | 0.795 | -log(1.2) | 0.015 |
| 0.84 | 0.898 | 0.816 | -log(1.1) | 0.027 |
| 0.84 | 0.786 | 0.864 | log(1.1) | 0.084 |
| 0.84 | 0.739 | 0.887 | log(1.2) | 0.129 |

Building on the last example, we consider adaptive Phase 2/3 oncology designs with dose optimization. Three hypothesis testing strategies are explored to address concerns about inconsistency, with an emphasis on controlling Type I error. One of the two doses in Phase 2 is selected to advance to Phase 3 based on early efficacy and safety data. The probability of picking-the-winner with greater survival benefit under the null hypothesis of no treatment effect at both doses is assumed to be $w$. It can take any value between 0 and 1, indicating a random high when it is greater than 0.5 and a random low when it is less than 0.5. Both can exacerbate the inconsistency issue due to regression to the mean. The probability depends on the dose selection criteria and the correlations of the endpoints involved in dose selection [6], and its estimation is out of the scope of this paper. Notably, when $w=0.5$ there is effectively no adaptive dose selection based on trial data. We further assume that patients treated at the selected dose and those from the control arm in Stage 1 contribute $t$ information fraction to the combined data from the two stages (total survival information = 510/4). The difference in the observed treatment effect (-logarithm of the hazard ratio) between Stage 1 and Stage 2 will be examined for consistency as measured by a cutoff point of $c$ (e.g., log(1.1) or log(1.2) based on Table 1).

We investigate the following three hypothesis testing strategies:

- Conservative: Include Stage 1 data in the analysis if the difference is less than $c$; otherwise, only include Stage 2 data in the analysis. For example, when $c=\log(1.15)$, data in row 3 will be combined but data in row 4 will not. This strategy may be preferred by a regulatory agency.
- Aggressive: Include Stage 1 data in the analysis if the difference is greater than $-c$; otherwise, only include Stage 2 data in the analysis. For example, when $c=\log(1.15)$, data in row 2 will be combined but data in row 1 will not. This strategy may be preferred by a trial sponsor.
- Neutral: Include Stage 1 data in the analysis if the absolute difference is less than $c$; otherwise, only include Stage 2 data in the analysis. This strategy provides a reasonable balance between the above two, as the first may be viewed as too conservative by trial sponsors while the second may be considered cherry-picking by regulatory agencies.

The primary analysis, with or without including the Stage 1 data, will be conducted at the adjusted alpha-level ($\alpha^*$) to control the overall Type error at $\alpha$. Details on how to find $\alpha^*$ are provided in the Appendix.

Figure 1 shows $\alpha^*$ values across various scenarios when $t$ is set at 30%. To maintain strong control of Type I error, we restrict $w$ to a range between 0.5 and 1.0 [6]. Clearly, little penalty needs to be paid under the conservative strategy, while a lot must be paid under the aggressive strategy. The penalty for the neutral strategy falls in between. As expected, the penalty increases with $c$ and $w$. Under the neutral strategy, $\alpha^*$ decreases slowly from 0.0204 at $w$=0.5 to 0.0183 at $w$=1.0 when $c$=log(1.1) and from 0.0188 at $w$=0.5 to 0.0160 at $w$=1.0 when $c$=log(1.2).

Figure 2 takes a closer look at $\alpha^*$ under the neutral strategy when $w$ is set at an extreme value of special interest (0.5 or 1.0). While $\alpha^*$ decreases as $t$ increases, as expected, it is not monotonic in $c$ and remains relative stable when $c$ is in a range of practical interest (e.g., log(1.1) to log(1.2)). In practice, patients treated at the unselected doses may have a different survival follow-up and treatment pattern compared to those treated at the selected dose, it is hard to know whether a true winner or loser has been picked based on the observed data. If the survival data is reliable, $w$ would be 1 if a winner is picked and 0 if a loser is picked (0 is technically changed to 0.5 in $\alpha^*$ calculation to maintain strong control of Type I error). If the survival data is unreliable, we would need to rely on an estimated $w$ to calculate $\alpha^*$. When the estimation is infeasible, we may need to conservatively set it at 1. However, if the overall efficacy and safety data are comparable between the two doses, as when the dose-response curve is flat between the two, it is reasonable to set it at 0.5. Aligned with the intuition, the resulting penalty is minimal, if any, when data between Stage 1 and Stage 2 are further consistent.

In this paper, we have addressed the concern about inconsistency of results in the Type I error control of adaptive Phase 2/3 designs. Choice of the cutoff point for measuring consistency not only impacts $\alpha^*$ but also study power, a topic under investigation. The introduction of $c$ under the neutral strategy results in a hybrid design: it becomes more inferential as $c$ increases and is more operational as $c$ decreases. To what extent the adaptive design is inferential or operational is data-driven and guided by $c$. It is increasingly realized that the assumption of $w$=1 under the prevailing statistical approaches is too conservative. While novel alternative approaches have been proposed [6-9], none have taken the inconsistency concern into consideration. This investigation paves the way for further research in a less explored area.

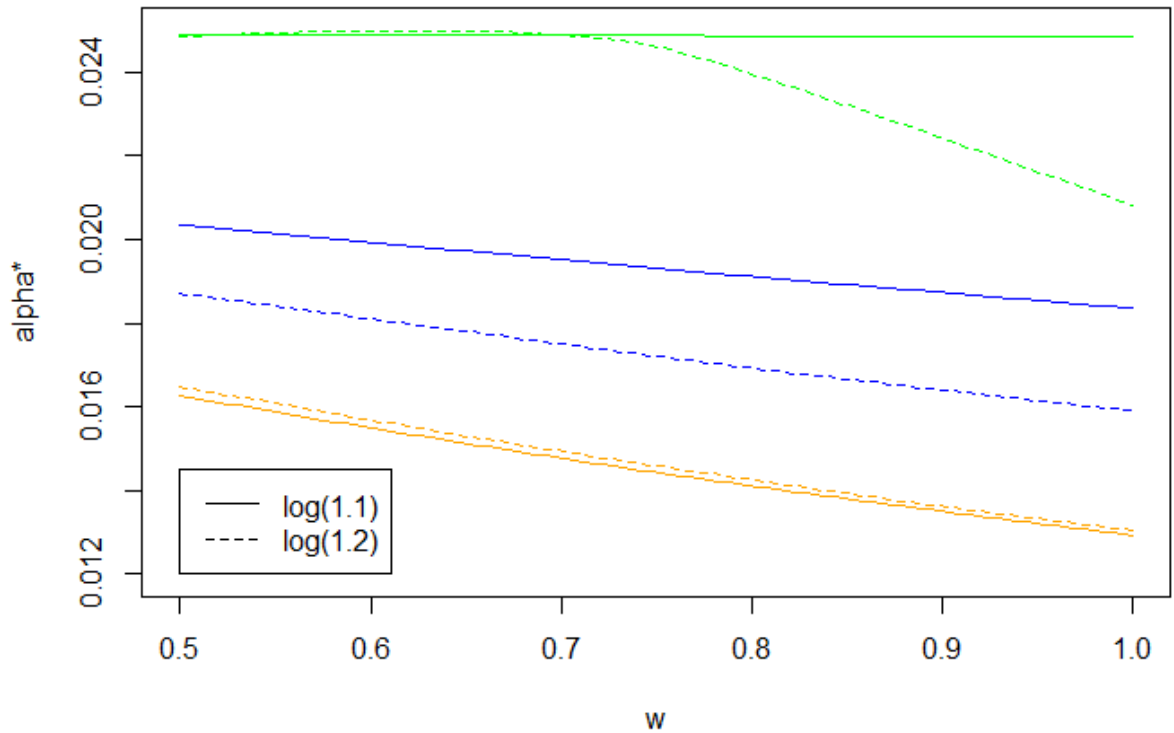

Figure 1. Plot of $\alpha^*$ as $w$ increases from 0.5 to 1.0 for the three statistical testing strategies (from top to bottom: green=conservative, blue=neutral, orange=aggressive) when $c$=log(1.1) or log (1.2).

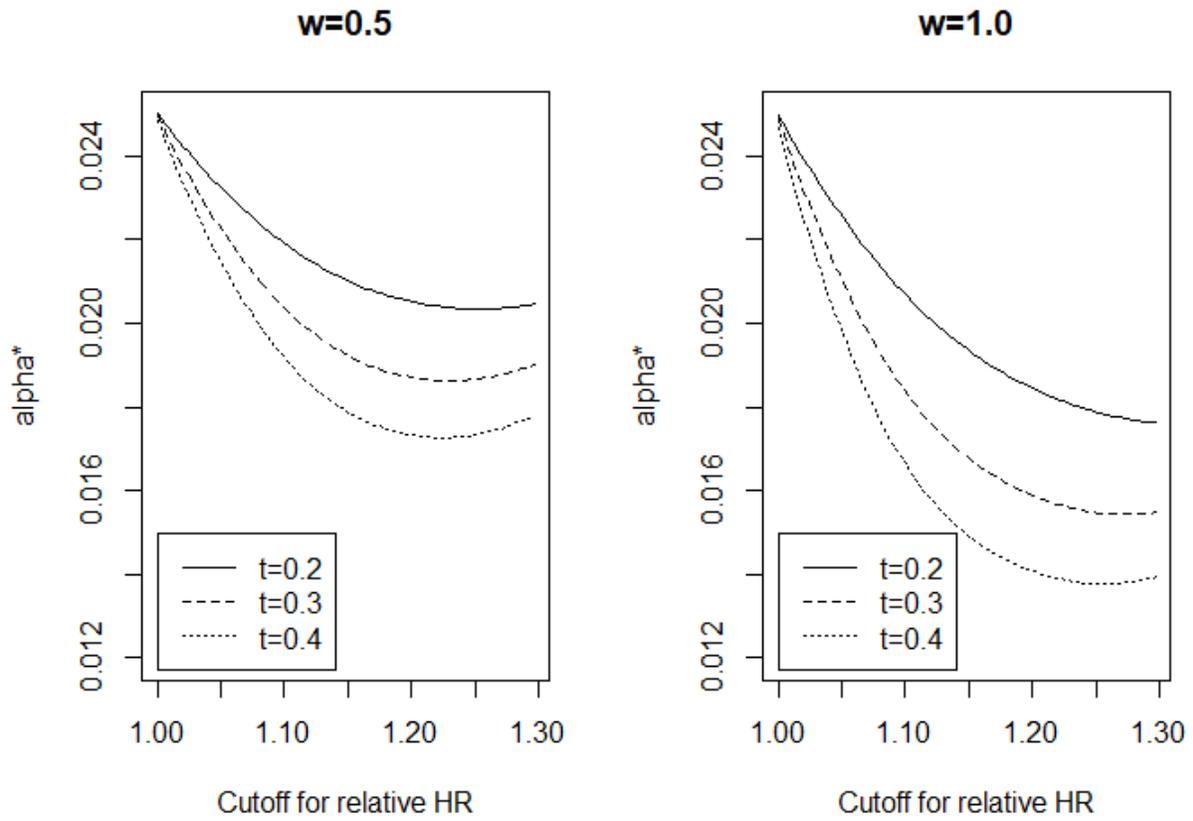

Figure 2: Plot of $\alpha^*$ under the neutral strategy as $c$ increases from $\log(1.0)$ to $\log(1.3)$ (left panel: $w$=0.5; right panel: $w$=1.0) when $t$=0.2, 0.3, or 0.4.

# Appendix: Calculation of $\alpha^*$ for overall Type I error control

The overall number of events between the selected dose and the control in the adaptive Phase 2/3 design is N, and the information unit is I=N/4. Let $Y_{1j}$ be the log-rank test of overall survival in Stage 1 patients at dose level $j$, $j = 1, 2$. Let $Y_{1s}$ be the corresponding log-rank test for the selected dose and $w = P(Y_{1s} = \max\{Y_{11}, Y_{12}\})$, i.e., probability of picking-the-winner due to random high. Let $Y_{2s}$ be the log-rank test of overall survival based on Stage 2 patients. The observed log hazard ratio is approximately $Y_{1s}/\sqrt{tI}$ at Stage 1 and $Y_{2s}/\sqrt{(1-t)I}$ at Stage 2. The log-rank test for the combined data can be approximated with $Y_s = \sqrt{t}Y_{1s} + \sqrt{1-t}Y_{2s}$, which inflates the Type I error rate when $w > 0.5$. The condition for combining the data for analysis is $\frac{Y_{2s}}{\sqrt{(1-t)I}} - \frac{Y_{1s}}{\sqrt{tI}} < c$ under the conservative strategy, $\frac{Y_{2s}}{\sqrt{(1-t)I}} - \frac{Y_{1s}}{\sqrt{tI}} > -c$ under the aggressive strategy, and $\left|\frac{Y_{1s}}{\sqrt{tI}} - \frac{Y_{2s}}{\sqrt{(1-t)I}}\right| < c$ under the neutral strategy. When the condition is met, $Y_s$ is used for hypothesis testing; otherwise, $Y_{2s}$ is used, with both tested at the same $\alpha^*$ level.

We only provide details under the neutral strategy for hypothesis testing. Derivations under the two other strategies are simpler. The overall Type I error under the neutral strategy composes of the following four components.

$$P\left(\left|\frac{Y_{1s}}{\sqrt{tI}} - \frac{Y_{2s}}{\sqrt{(1-t)I}}\right| < c, \sqrt{t}Y_{1s} + \sqrt{1-t}Y_{2s} > Z_{1-\alpha^*} | Y_{1s} = \max\{Y_{11}, Y_{12}\}\right) w \stackrel{\text{def}}{=} Aw$$

$$P\left(\left|\frac{Y_{1s}}{\sqrt{tI}} - \frac{Y_{2s}}{\sqrt{(1-t)I}}\right| < c, \sqrt{t}Y_{1s} + \sqrt{1-t}Y_{2s} > Z_{1-\alpha^*} | Y_{1s} = \min\{Y_{11}, Y_{12}\}\right) (1-w) \stackrel{\text{def}}{=} B(1-w)$$

$$P\left(\left|\frac{Y_{1s}}{\sqrt{tI}} - \frac{Y_{2s}}{\sqrt{(1-t)I}}\right| > c, Y_{2s} > Z_{1-\alpha^*} | Y_{1s} = \max\{Y_{11}, Y_{12}\}\right) w \stackrel{\text{def}}{=} Cw$$

$$P\left(\left|\frac{Y_{1s}}{\sqrt{tI}} - \frac{Y_{2s}}{\sqrt{(1-t)I}}\right| > c, Y_{2s} > Z_{1-\alpha^*} | Y_{1s} = \min\{Y_{11}, Y_{12}\}\right) (1-w) \stackrel{\text{def}}{=} D(1-w)$$

We rewrite each of them as a summation of the probability distribution function of multivariate normal variables for computational ease. The details for A are provided below. The details for others are omitted due to similarity.

$$A = P\left(\left|\frac{\max\{Y_{11}, Y_{12}\}}{\sqrt{tI}} - \frac{Y_{2s}}{\sqrt{(1-t)I}}\right| < c\right)$$
$$-P\left(\left|\frac{\max\{Y_{11}, Y_{12}\}}{\sqrt{tI}} - \frac{Y_{2s}}{\sqrt{(1-t)I}}\right| < c, \sqrt{t}\max\{Y_{11},Y_{12}\} + \sqrt{1-t}Y_{2s} < Z_{1-\alpha^*}\right)$$
$$= P\left(\frac{\max\{Y_{11}, Y_{12}\}}{\sqrt{tI}} - \frac{Y_{2s}}{\sqrt{(1-t)I}} < c\right) - P\left(\frac{\max\{Y_{11}, Y_{12}\}}{\sqrt{tI}} - \frac{Y_{2s}}{\sqrt{(1-t)I}} < -c\right)$$
$$-P\left(\frac{\max\{Y_{11}, Y_{12}\}}{\sqrt{tI}} - \frac{Y_{2s}}{\sqrt{(1-t)I}} < c, \sqrt{t}\max\{Y_{11},Y_{12}\} + \sqrt{1-t}Y_{2s} < Z_{1-\alpha^*}\right)$$
$$+P\left(\frac{\max\{Y_{11}, Y_{12}\}}{\sqrt{tI}} - \frac{Y_{2s}}{\sqrt{(1-t)I}} < -c, \sqrt{t}\max\{Y_{11},Y_{12}\} + \sqrt{1-t}Y_{2s} < Z_{1-\alpha^*}\right)$$
$$= P\left(\frac{Y_{11}}{\sqrt{tI}} - \frac{Y_{2s}}{\sqrt{(1-t)I}} < c, \frac{Y_{12}}{\sqrt{tI}} - \frac{Y_{2s}}{\sqrt{(1-t)I}} < c\right) - P\left(\frac{Y_{11}}{\sqrt{tI}} - \frac{Y_{2s}}{\sqrt{(1-t)I}} < -c, \frac{Y_{12}}{\sqrt{tI}} - \frac{Y_{2s}}{\sqrt{(1-t)I}} < -c\right)$$
$$-P\left(\frac{Y_{11}}{\sqrt{tI}} - \frac{Y_{2s}}{\sqrt{(1-t)I}} < c, \frac{Y_{12}}{\sqrt{tI}} - \frac{Y_{2s}}{\sqrt{(1-t)I}} < c, \sqrt{t}Y_{11} + \sqrt{1-t}Y_{2s} < Z_{1-\alpha^*}, \sqrt{t}Y_{12} + \sqrt{1-t}Y_{2s} < Z_{1-\alpha^*}\right)$$
$$+P\left(\frac{Y_{11}}{\sqrt{tI}} - \frac{Y_{2s}}{\sqrt{(1-t)I}} < -c, \frac{Y_{12}}{\sqrt{tI}} - \frac{Y_{2s}}{\sqrt{(1-t)I}} < -c, \sqrt{t}Y_{11} + \sqrt{1-t}Y_{2s} < Z_{1-\alpha^*}, \sqrt{t}Y_{12} + \sqrt{1-t}Y_{2s} < Z_{1-\alpha^*}\right)$$

$$B = P\left(\frac{Y_{11}}{\sqrt{tI}} - \frac{Y_{2s}}{\sqrt{(1-t)I}} > -c, \frac{Y_{12}}{\sqrt{tI}} - \frac{Y_{2s}}{\sqrt{(1-t)I}} > -c, \sqrt{t}Y_{11} + \sqrt{1-t}Y_{2s} > Z_{1-\alpha^*}, \sqrt{t}Y_{12} + \sqrt{1-t}Y_{2s} > Z_{1-\alpha^*}\right)$$
$$-P\left(\frac{Y_{11}}{\sqrt{tI}} - \frac{Y_{2s}}{\sqrt{(1-t)I}} > c, \frac{Y_{12}}{\sqrt{tI}} - \frac{Y_{2s}}{\sqrt{(1-t)I}} > c, \sqrt{t}Y_{11} + \sqrt{1-t}Y_{2s} > Z_{1-\alpha^*}, \sqrt{t}Y_{12} + \sqrt{1-t}Y_{2s} > Z_{1-\alpha^*}\right)$$

$$C = P\left(\frac{Y_{11}}{\sqrt{tI}} - \frac{Y_{2s}}{\sqrt{(1-t)I}} < -c, \frac{Y_{12}}{\sqrt{tI}} - \frac{Y_{2s}}{\sqrt{(1-t)I}} < -c, Y_{2s} > Z_{1-\alpha^*}\right) + P(Y_{2s} > Z_{1-\alpha^*})$$
$$-P\left(\frac{Y_{11}}{\sqrt{tI}} - \frac{Y_{2s}}{\sqrt{(1-t)I}} < c, \frac{Y_{12}}{\sqrt{tI}} - \frac{Y_{2s}}{\sqrt{(1-t)I}} < c, Y_{2s} > Z_{1-\alpha^*}\right)$$
$$= P\left(\frac{Y_{11}}{\sqrt{tI}} - \frac{Y_{2s}}{\sqrt{(1-t)I}} < -c, \frac{Y_{12}}{\sqrt{tI}} - \frac{Y_{2s}}{\sqrt{(1-t)I}} < -c, Y_{2s} > Z_{1-\alpha^*}\right) + \alpha^*$$
$$-P\left(\frac{Y_{11}}{\sqrt{tI}} - \frac{Y_{2s}}{\sqrt{(1-t)I}} < c, \frac{Y_{12}}{\sqrt{tI}} - \frac{Y_{2s}}{\sqrt{(1-t)I}} < c, Y_{2s} > Z_{1-\alpha^*}\right)$$

$$D = P\left(\frac{Y_{11}}{\sqrt{tI}} - \frac{Y_{2s}}{\sqrt{(1-t)I}} > c, \frac{Y_{12}}{\sqrt{tI}} - \frac{Y_{2s}}{\sqrt{(1-t)I}} > c, Y_{2s} > Z_{1-\alpha^*}\right) + \alpha^*$$
$$-P\left(\frac{Y_{11}}{\sqrt{tI}} - \frac{Y_{2s}}{\sqrt{(1-t)I}} > -c, \frac{Y_{12}}{\sqrt{tI}} - \frac{Y_{2s}}{\sqrt{(1-t)I}} > -c, Y_{2s} > Z_{1-\alpha^*}\right)$$

Notice that $Y_{11}$, $Y_{12}$, and $Y_{2s}$ are all standard normal variables. The correlation between $Y_{11}$ and $Y_{12}$ is 0.5 due to the sharing the control arm, and both are independent of $Y_{2s}$. The variance-covariance structure of the variables involved in each probability function can be derived accordingly. The R code for $\alpha^*$ calculation under the neutral strategy is provided below. A sample code "solve_astar(0.025, c = log(1.1), t = 0.3, I = 510/4, w = 0.6)" returns $\alpha^*$=0.0199 that controls overall alpha at 0.025 when $c = \log(1.1)$, $t = 0.3$, $I = 510/4$, and $w = 0.6$.

```
library(mvtnorm)

eq1 <- function(c, t, I, astar) {
  s1 <- matrix(c(1/(t*I)+1/((1-t)*I), 1/(2*t*I)+1/((1-t)*I),
            1/(2*t*I)+1/((1-t)*I), 1/(t*I)+1/((1-t)*I)), 2, 2)
  s2a <- s1
  s2b <- matrix(c(0, -1/(2*sqrt(I)), -1/(2*sqrt(I)), 0), 2, 2)
  s2c <- matrix(c(1, 1-t/2, 1-t/2, 1), 2, 2)
  s2 <- rbind(cbind(s2a, s2b), cbind(s2b, s2c))
  f11 <- pmvnorm(upper = c(c, c), sigma = s1)[1]
  f12 <- pmvnorm(upper = c(-c, -c), sigma = s1)[1]
  f13 <- pmvnorm(upper = c(c, c, qnorm(1-astar), qnorm(1-astar)),
            sigma = s2)[1]
  f14 <- pmvnorm(upper = c(-c, -c, qnorm(1-astar), qnorm(1-astar)),
            sigma = s2)[1]
  f11 - f12 - f13 + f14
}

eq2 <- function(c, t, I, astar) {
  s1 <- matrix(c(1/(t*I)+1/((1-t)*I), 1/(2*t*I)+1/((1-t)*I),
            1/(2*t*I)+1/((1-t)*I), 1/(t*I)+1/((1-t)*I)), 2, 2)
  s2a <- s1
  s2b <- matrix(c(0, -1/(2*sqrt(I)), -1/(2*sqrt(I)), 0), 2, 2)
  s2c <- matrix(c(1, 1-t/2, 1-t/2, 1), 2, 2)
  s2 <- rbind(cbind(s2a, s2b), cbind(s2b, s2c))
  f21 <- pmvnorm(lower = c(-c, -c, qnorm(1-astar), qnorm(1-astar)),
            sigma = s2)[1]
  f22 <- pmvnorm(lower = c(c, c, qnorm(1-astar), qnorm(1-astar)),
            sigma = s2)[1]
  f21 - f22
}

eq3 <- function(c, t, I, astar) {
```

```r
    s1 <- matrix(c(1/(t*I)+1/((1-t)*I), 1/(2*t*I)+1/((1-t)*I),
                   1/(2*t*I)+1/((1-t)*I), 1/(t*I)+1/((1-t)*I)), 2, 2)
    s3 <- rbind(cbind(s1, rep(-1/sqrt((1-t)*I), 2)),
                c(rep(-1/sqrt((1-t)*I), 2), 1))
    f31 <- astar
    f32 <- pmvnorm(upper = c(-c, -c, Inf), lower = c(-Inf, -Inf, qnorm(1-astar)),
                   sigma = s3)[1]
    f33 <- pmvnorm(upper = c(c, c, Inf), lower = c(-Inf, -Inf, qnorm(1-astar)),
                   sigma = s3)[1]
    f31 + f32 - f33
}

eq4 <- function(c, t, I, astar) {
    s1 <- matrix(c(1/(t*I)+1/((1-t)*I), 1/(2*t*I)+1/((1-t)*I),
                   1/(2*t*I)+1/((1-t)*I), 1/(t*I)+1/((1-t)*I)), 2, 2)
    s3 <- rbind(cbind(s1, rep(-1/sqrt((1-t)*I), 2)),
                c(rep(-1/sqrt((1-t)*I), 2), 1))
    f41 <- astar
    f42 <- pmvnorm(lower = c(c, c, qnorm(1-astar)),
                   sigma = s3)[1]
    f43 <- pmvnorm(lower = c(-c, -c, qnorm(1-astar)),
                   sigma = s3)[1]
    f41 + f42 - f43
}

calc_typeI <- function(astar, c, t, I, w) {
  eq1(c = c, t = t, I = I, astar = astar)*w +
    eq2(c = c, t = t, I = I, astar = astar)*(1-w) +
    eq3(c = c, t = t, I = I, astar = astar)*w +
    eq4(c = c, t = t, I = I, astar = astar)*(1-w)
}

solve_astar <- function(alpha, c, t, I, w) {
  op_fn <- function(astar, alpha, c, t, I, w) {
    (calc_typeI(astar, c, t, I, w) - alpha)^2
  }
  optimize(op_fn, lower = 0, upper = alpha, alpha = alpha, c = c, t = t, I = I,
           w = w)$minimum
}
```